\documentclass{article}

\usepackage{graphicx}
\graphicspath{{Figures/}}

\usepackage{subfig}

\usepackage{booktabs}

\usepackage{tabu}

\usepackage{xspace}

\usepackage{verbatim}

\newcommand{\Bold}[1]{\textbf{#1}}

\newcommand{\Example}[1]{\textit{\footnotesize #1}}

\title{Group-Level Graph Visualization Taxonomy}
\author{Bahador Saket, Paolo Simonetto and Stephen Kobourov\\Computer Science Department, University of Arizona}
\date{}

\begin{document}
\maketitle

\begin{abstract}
Task taxonomies for graph and network visualizations focus on tasks commonly encountered when analyzing graph connectivity and topology. However, in many application fields such as the social sciences (social networks), biology (protein interaction models), software engineering (program call graphs), connectivity and topology information is intertwined with group, clustering, and hierarchical information. Several recent visualization techniques, such as BubbleSets, LineSets and GMap, make explicit use of grouping and clustering, but evaluating such visualization has been difficult due to the lack of standardized group-level tasks. With this in mind, our goal is to define a new set of tasks that assess group-level comprehension. We propose several types of group-level tasks and provide several examples of each type. Finally, we characterize some of the proposed tasks using the multi-level typology of abstract visualization tasks. We believe that adding group-level tasks to the task taxonomy for graph visualization would make the taxonomy more useful for the recent graph visualization techniques. It would help evaluators define and categorize new tasks, and it would help generalize individual results collected in controlled experiments.

\end{abstract}

\section{Introduction} \label{SEC:Introduction}
Graphs are used to describe a set of entities (nodes) and their relationships (edges). Graphs and networks (used interchangeably here) are typically visualized using a node-link diagram, where nodes are depicted as points, and edges as line segments connecting the corresponding points.

Several studies have tested the readability of node-link diagrams. In particular, Purchase et al.~\cite{Graph_layout_evaluation_Purchase,Graph_layout_evaluation_UML} examined how graph drawing aesthetics such as edge crossings and display of symmetries impact performance of graph reading tasks, such as path tracing. Huang et al.~\cite{IV_Huang1, IV_Huang2, IV_Huang3, IV_Huang4} used eye-tracking and control experiments to understand visual network perception. Archambault, Ghani and Farrugia evaluated perceptual characteristics and memorability in dynamic, animated graphs~\cite{IV_ghani2, IV_Farrugia, IV_Arcambault1, IV_Arcambault2}.

The results of these studies are difficult to compare, because of the absence of a standardized approach to graph evaluation studies. Seemingly non-influential decisions, such as the choice or phrasing of the tasks, may have a significant impact on the results. In an attempt to mitigate this problem, visual data analysis tasks can be organized and categorized in taxonomies. Brehmer and Munzner~\cite{IV_Brehmer} reviewed and compared a large number of earlier studies, and in doing so provided a schema that blends with the existing taxonomies and allow to fully characterize motivation, methods, and input/output information for each task. 

\begin{figure}[t]
\centering \hfill
\subfloat[]{\includegraphics[width=0.23\columnwidth]{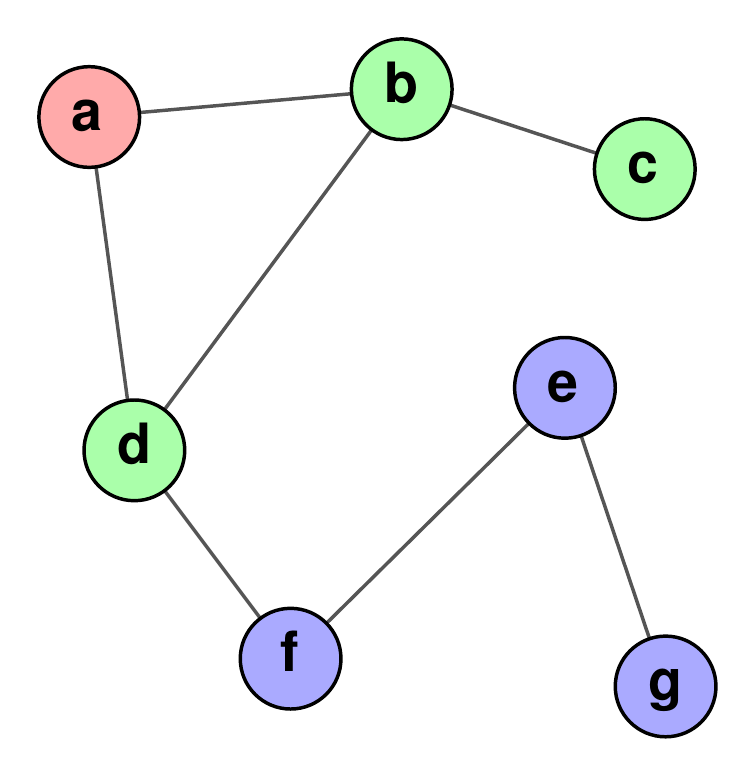}} \hfill
\subfloat[]{\includegraphics[width=0.23\columnwidth]{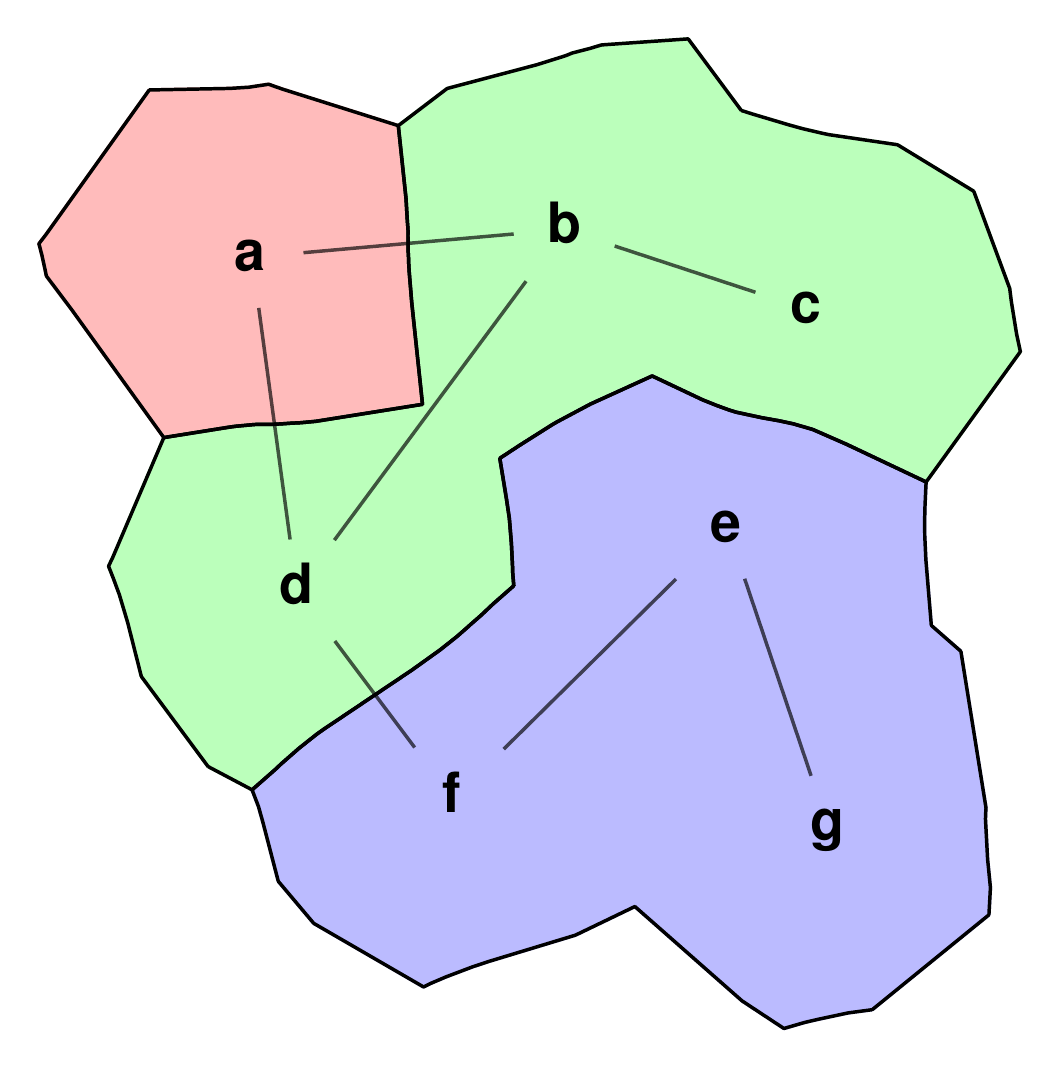}} \hfill
\subfloat[]{\includegraphics[width=0.23\columnwidth]{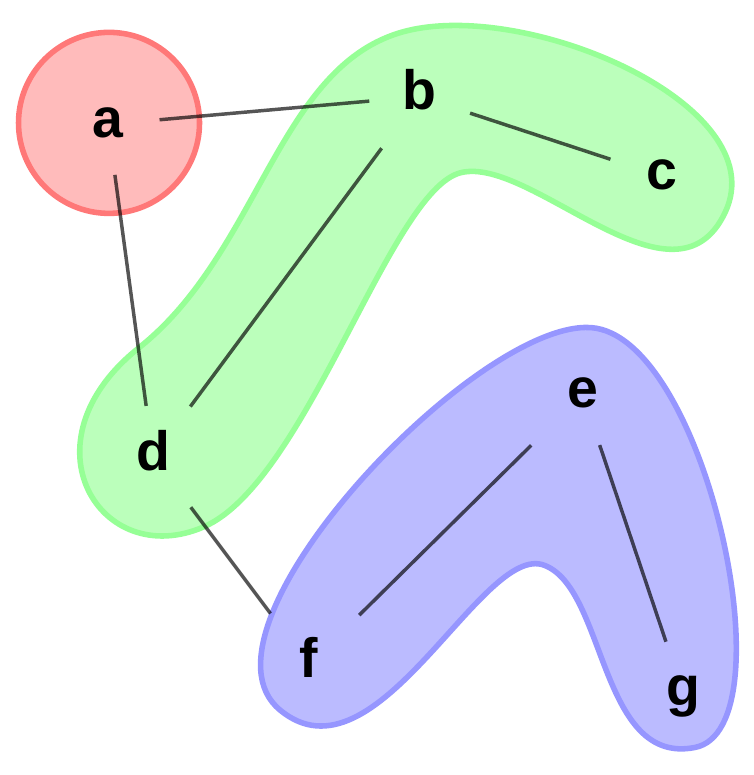}} \hfill
\subfloat[]{\includegraphics[width=0.23\columnwidth]{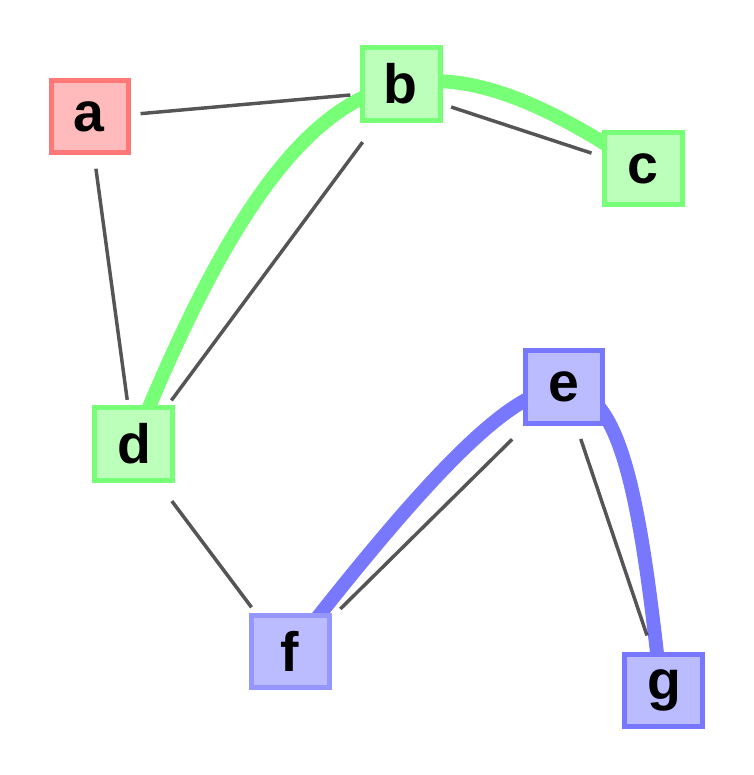}} \hfill
\caption{Four different methods for representing groups of nodes in a network. \textbf{a)} in node-link diagrams, groups are often encoded with node colors. \textbf{b)} in GMap, groups are depicted so that they resemble countries of a geographical maps. \textbf{c)} in BubbleSets, isocontours are drawn to enclose the nodes belonging to same set. \textbf{d)} in LineSets, elements have of the same set are identified by colored labels and links.}
\label{FIG:prop-p3t}
\end{figure}

Although task taxonomies are available for a broad range of visualization techniques, including node-link diagrams~\cite{IV_BL}, none specifically deal with the visualization of clustered graphs. Clustered graphs, i.e.,   graphs where nodes are grouped based on a priori knowledge or structural properties, are fairly common (for example, communities in social networks~\cite{IV_Mishra}, or co-activated proteins in protein-protein interaction networks~\cite{IV_Barsky}). Clustered graph visualization is common in many different domains: from  information spatialization~\cite{sf-sm-03,fabrikant06}, to self organizing maps of documents, coupled with geographic information systems~\cite{Sk02}, to general maps of science showing groups of scientific disciplines~\cite{Boyack05}, to maps of computer science~\cite{mocs}. It has been shown that augmenting node-link diagrams with spatial features can improve graph revisitation tasks~\cite{Ghani2011}.
This is used in 
visualizations that explicitly draw boundaries to indicate the grouping: BubbleSets~\cite{IV_Collins},
LineSets~\cite{IV_Alper}, and GMap~\cite{IV_Gansner}; see Fig.~\ref{FIG:prop-p3t}.

We aim to expand the standard graph tasks taxonomy by providing a set of tasks relevant to the analysis of clustered graphs. We will classify these tasks according the kind of information required to solve them, and we will provide examples on how to fully describe them according to the multi-level typology of abstract visualization tasks proposed by Brehmer and Munzner~\cite{IV_Brehmer}.

As some readers might be unfamiliar with the context that motivates this work, we begin with a brief review of existing task taxonomies for graph visualizations, and the multi-level typology of abstract visualization tasks. In Section 3 we augment the task taxonomy for graph visualization by introducing a new set of tasks related to groups of nodes (Group-Level Tasks) and we characterize a couple of the newly proposed tasks using the multi-level typology of abstract visualization tasks. In Section 4 we discuss the potential value of the proposed group-level task taxonomies for graph visualization and show how these group-level tasks can be used in future evaluation studies.

\section{Task Taxonomies for Graph Visualization}
Brehmer and Munzner~\cite{IV_Brehmer} organize the vast previous work on task taxonomies in visualization, highlighting their advantages and disadvantages. They point out as the major shortcoming of most approaches, the lack of a global view of the task: high-level categories often ignore how the tasks are performed, while low level categories often ignore why the tasks are performed. In order to close this gap, they develop a multi-level typology that helps create a complete description of a task.

\begin{figure}[t]
\centering
\includegraphics[width=\columnwidth]{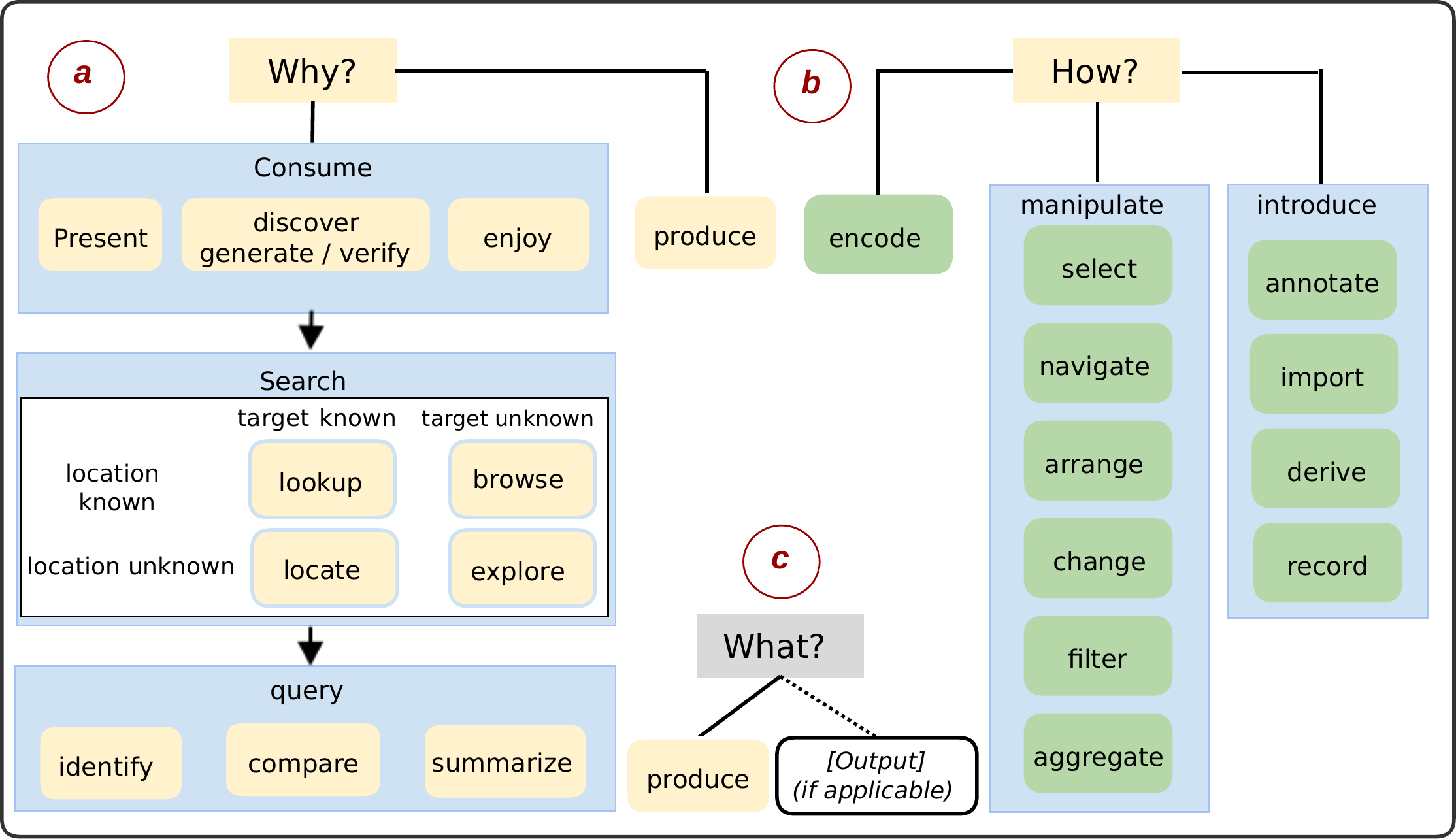}
\caption{Multi-level typology of abstract visualization tasks. The typology spans \textsc{Why, how} and \textsc{what}. Figure from~\cite{IV_Brehmer} used with permission.} \label{FIG:Munzner}
\end{figure}

This multi-level typology encompasses three main questions: \textsc{Why, how} and \textsc{what}. The \textsc{Why} part of the typology allows us to describe why a task is performed, includes multiple levels of specificity, and a narrowing of scope from high-level (consume vs. produce) to mid-level (search) to low-level (query); see Fig.~\ref{FIG:Munzner}a. The \textsc{how} part of the typology allows us to describe how a task is performed, and this part includes three classes of methods: those for encoding data, those for manipulating existing elements in a visualization, and those for introducing new elements into a visualization; see Fig.~\ref{FIG:Munzner}b. Finally, the \textsc{what} part of the typology allows us to describe what are the inputs and outputs for a given task; see Fig.~\ref{FIG:Munzner}c. This definition is purely abstract and enables the translation of any type of relevant task into the why/how/what framework, making it clear and almost ready for implementation. 

The work of Brehmer and Munzner, however, is not meant to replace model-oriented taxonomies, but rather to ``encompass and complement these specific classification systems''. In fact, topic-specific taxonomies provide details about low-level tasks, that are necessarily abstracted in the very general approach of Brehmer and Munzner. Instead, they provide the tools to put these low level tasks in context, guiding the evaluation designer in providing information 
such as user expertise and motivation. 

A couple of existing taxonomies served as foundations for our work. Amar et al.~\cite{IV_RA1} describe a set of ten primitive analysis task types, representative of the kinds of specific questions that one may ask when working with tabular data (e.g., Retrieve Value, Find Extremum, etc.).  The task taxonomy for graph visualization developed by Lee et al.~\cite{IV_BL} is built upon these tasks, but the authors found that it was necessary to define additional low-level tasks, such as scan and set operations, and the graph-specific low-level task: ``find adjacent nodes''. The final set of tasks was then organized into four groups: topology-based tasks, attribute-based tasks, browsing tasks, and overview tasks.

\begin{table}[p]
  \small
  \setlength{\tabulinesep}{7pt}
  \begin{tabu}{X}
    \toprule
    \Bold{Group Only Tasks} \\
    Find the set of group-neighbors of a given a group. ---
    How many groups are neighbors of a given group? ---
    Which group has the maximum (minimum) number of neighboring groups? ---
    Find the set of groups accessible from a group. ---
    How many groups are accessible from a group? ---
    Find the set of groups one group away from a given group. ---
    Given two groups, find a set of groups that are adjacent to both of them. ---
    Find the shortest path between two groups. ---
    Find a group with specific characteristics (e.g., red background). ---
    Find the group with largest (smallest) area. ---
    Are the given two groups neighbors? ---
    Find a group whose removal from the visualization disconnects the map. ---
    How many groups are there? ---
    Find a group which has the longest (shortest) boundary with a given group. \\
    \midrule

    \Bold{Group-Node Tasks} \\
    Given a node X (with specific characteristics), find the group which contains X. ---
    Count number of nodes in a specific group. ---
    Find the group with the maximum (minimum) number of nodes. ---
    Given two nodes X and Y, check whether these two nodes belong to the same group. ---
    List groups which contain nodes with specific characteristics. \\
    \midrule

    \Bold{Group-Link Tasks}  \\
    Count number of links in a given group. ---
    Find the group with the maximum (minimum) number of links. ---
    Find the most sparsely (most densely) connected group. ---
    Find the group that contains the longest link (or the pair of groups at the endpoints of the longest link). ---
    List groups which contain a link with specific characteristics (e.g., longest, heaviest). \\
    \midrule
    
    \Bold{Group-Network Tasks} \\
    Find two groups with a link between them, whose removal disconnects the network. ---
    Given two groups, can they be disconnected by removing no more than $n$ links? ---
    Find a group which has the node with highest (lowest) degree. ---
    Find the path X-Y-Z; are nodes X and Z in the same group? ---
    Given two nodes in different groups A and B, what is the smallest number of groups that need to be visited on a path from A to B? \\
    \bottomrule 
  \end{tabu}
\caption{Group Level Tasks} \label{TAB:tasks}
\end{table}

\section{Task Taxonomy for Clustered Graph Visualization}
We defined a list of 29 group-level tasks both from studying the user interaction with visualizations such as BubbleSets~\cite{IV_Collins}, LineSets~\cite{IV_Alper} and GMap~\cite{IV_Gansner}, and from interviewing experts in the field. We divided these group-level tasks into four subcategories according to the information required to solve them. 

\begin{itemize}
\item \Bold{Group Only Tasks}:  Tasks in this category can be performed by only considering the groups, so that no node or edge information is required. \Example{\Bold{For example:} Given a group X, find all groups neighbors of group X.}
\item \Bold{Group-Node Tasks}: Tasks in this category can be performed by only considering group and node information. \Example{\Bold{For example:} Find the group with the maximum number of nodes.}
\item \Bold{Group-Link Tasks}: Tasks in this category can be performed by only considering group and edge information. \Example{\Bold{For example:} Find a group with the minimum number of links.}
\item \Bold{Group-Network Tasks}: Tasks in this category can be performed by only considering group, node and edge information. \Example{\Bold{For example:} Find a group which has the node with the highest degree.}
\end{itemize}

\begin{table}[p]
   \small
   \setlength{\tabulinesep}{7pt}
   \begin{tabu}{X[0.4]X[0.95]X[0.5]X[0.4]}
     \toprule  
     Tasks & \textsc{Why} & \textsc{What} & \textsc{how} \\
     \midrule
     Given two groups X and Y, are these two groups neighbors? & The purpose of the task is to discover whether groups X and Y are adjacent. The targets are known (X and Y are specified). If the participants are aware of the location of these two groups then this is a Lookup task; otherwise this is a Locate task; see Fig.~\ref{FIG:Munzner}a). 
     Once the participants find both groups, they need to identify whether they are adjacent in the map.
     & The input for the task are groups X and Y. The output is Yes if two groups are neighbors and No otherwise; see Fig.~\ref{FIG:Munzner}c).
     & The participant needs to be able to tell groups X and Y apart and to check whether they have a common boundary or another group is in between.  \\
     & Discover + (Look up + Locate) + Identify &
     Input: Groups X and Y \newline Output: Yes / No &
     Select \\
     \midrule
      
     Find the 3 groups with the most links. & The purpose of the task is to discover which 3 groups have the most links. Since the group characteristics (e.g., name or color) are not given, the targets are unknown. The participant searches for the groups not knowing their location; this is an Explore task. Finally, the participant needs to produce the identity of the 3 groups. 
     & The input for the task is all the groups, including their nodes and links. The output is identity of the 3 groups with the most links.   
     & The participant needs to count (estimate) the number of links for each group and keep track of the largest ones. \\
     & Discover + Explore + Summarize &
     Input: Entire map \newline Output: Three groups &
     Derive + Select \\
     \bottomrule
  \end{tabu}
\caption{Examples of group-level tasks described using the multi-level typology of abstract visualization tasks} \label{TAB:link}
\end{table}

In table~\ref{TAB:tasks} we collected several group-level tasks divided by category. Each task can be used as-is, or can be combined in a macro task. The list cannot be exhaustive, meaning that is surely does not cover all the tasks of a given type. However, the tasks contained can serve as an inspiration both for the definition of new tasks, and for the definition of more specific taxonomies. For example Jianu et al.~\cite{IV_Radu} defined the tasks used in their evaluation based on the taxonomy of Lee et al.~\cite{IV_BL}, which has been constructed on the more primitive one of Amar et al.~\cite{IV_RA1}.

Moreover, this taxonomy offers other two advantages: 1) task taxonomies can help us to be aware of the most of the possible tasks that can be performed when analyzing specific visualization. 2) asking users to perform different categories of tasks (e.g., Topology-Based Tasks, Browsing Tasks~\cite{IV_BL}) especially in user performance evaluations can help authors to ensure that the spectrum of tasks is effectively covered.

\subsection{Relationship to Graph-Level Tasks}
\begin{figure}[t]
\centering \hfill
\subfloat[]{\includegraphics[width=0.40\columnwidth]{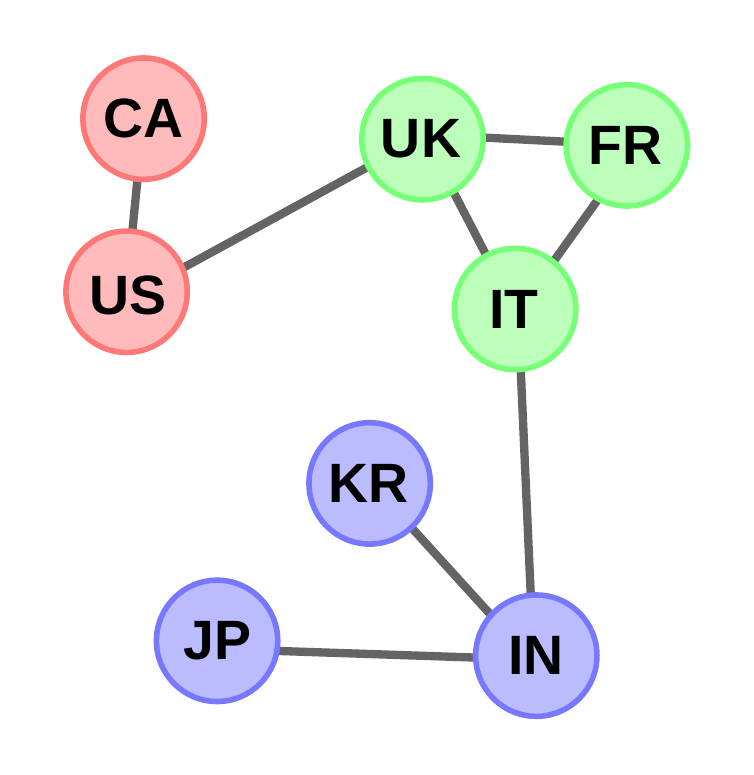}} \hfill
\subfloat[]{\includegraphics[width=0.40\columnwidth]{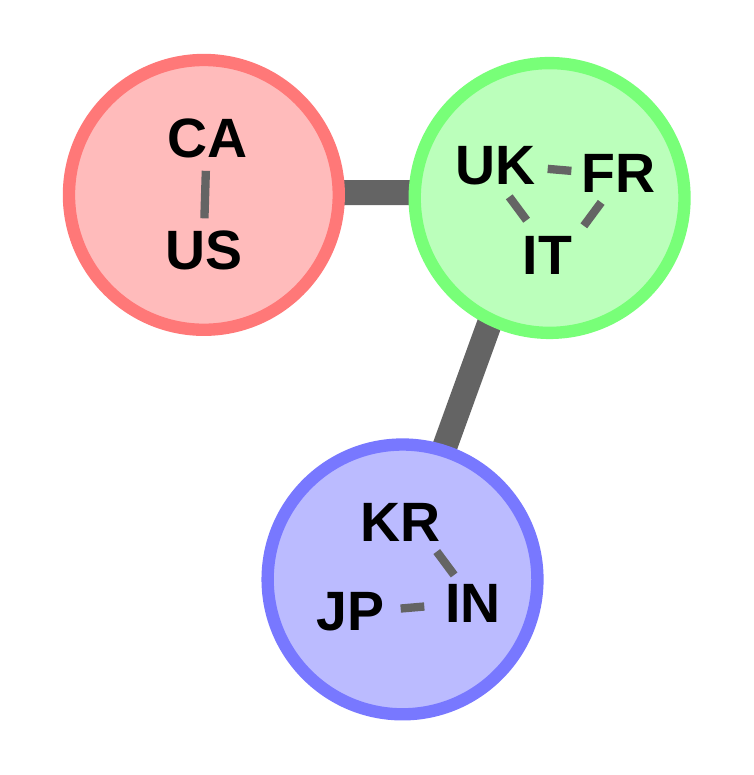}} \hfill
\caption{Construction of a metagraph. \textbf{a)} the original clustered graph. \Bold{b)} the metagraph. }
\label{FIG:metagraph}
\end{figure}

Many group-level tasks can be deduced as an extension of graph-related tasks. In fact, we can consider the clusters as metanodes, define metaedges according to the presence or absence of connections between the original nodes, and execute graph tasks on the metagraph generated by this process. Many graph characteristics, such as degree, adjacency or centrality of a node directly translate into relevant group properties. For instance, consider the graph of the connections of a courier company in Figure~\ref{FIG:metagraph}. We can analyze the graph to deduce that the best way to ship a parcel from the US to India require a transit through UK and Italy, but we can also analyze the metagraph to see that all parcels form North-America to Asia need to transit trough Europe.

This metagraph can be built in several different ways, leading to different insights on the data. For example, in representations where the groups share their boundaries (e.g. GMap), the metagraph can be built based on the group contacts. This would be useful, for instance, to identify the groups that have stronger interconnections with each other.

\subsection{Examples of Task Description}
In Table~\ref{TAB:link}, we provide a couple of examples of how our tasks can be described according to the typology of abstract visualization tasks~\cite{IV_Brehmer}. We can only provide examples rather than a full description in terms of \textsc{Why, how} and \textsc{what} because this requires information unavailable to us and known only by the evaluation designer. For example, participant who are asked to find a node X might or might not be aware of their approximate position, or might have different motivations for executing the task.

\section{Conclusions and Future Work}
Our primary contribution is in defining a taxonomy of group-level tasks. These tasks can be used to design and describe the operations that can be performed on a clustered graph, and should be used in conjunction with the work of Brehmer and Munzner to fully characterize an evaluation study. We showed this process using two examples.
We hope to use this taxonomy for two in-depth studies (in one we attempt to measure the role of explicit cluster boundaries when identifying groups of nodes, and in the other, the effectiveness of point-cloud, node-link, and map-based visualization).

The current taxonomy can be extended by expanding the set of defined tasks, or by further specializing it. We assumed that we are dealing with a non-overlapping clustering. However, overlapping groups are common in many applications such as social sciences (where one can participate in multiple communities). Incorporating overlapping groups would likely necessitate additional tasks to deal with the increased the number of possible operations and the increased drawing complexity.




\bibliographystyle{plain}
\bibliography{Bibliography}

\begin{thebibliography}{10}

\bibitem{IV_Alper}
B.~Alper, N.~Riche, G.~Ramos, and M.~Czerwinski.
\newblock Design study of linesets, a novel set visualization technique.
\newblock In {\em IEEE Trans. Visualization and Computer Graphics (TVCG)},
  pages 2259--2267, 2011.

\bibitem{IV_RA1}
R.~Amar, J.~Eagan, and J.~T. Stasko.
\newblock Low-level components of analytic activity in information
  visualization.
\newblock In {\em Symposium on Information Visualization (InfoVis '05)}, pages
  111--117, 2005.

\bibitem{IV_Arcambault1}
D~Archambault and H.~C. Purchase.
\newblock The mental map and memorability in dynamic graphs.
\newblock In {\em Pacific Visualization Symposium (PacificVis)}, pages 89--96,
  2012.

\bibitem{IV_Arcambault2}
D.~Archambault and H.~C. Purchase.
\newblock Mental map preservation helps user orientation in dynamic graphs.
\newblock In {\em Graph Drawing}, pages 475--486, 2013.

\bibitem{IV_Barsky}
A.~Barsky, T.~Munzner, Gardy. J., and R.~Kincaid.
\newblock Cerebral: Visualizing multiple experimental conditions on a graph
  with biological context.
\newblock In {\em Visualization and Computer Graphics}, pages 1253--1260, 2008.

\bibitem{Boyack05}
Kevin~W. Boyack, Richard Klavans, and Katy B{\"o}rner.
\newblock Mapping the backbone of science.
\newblock {\em Scientometrics}, 64:351--374, 2005.

\bibitem{IV_Brehmer}
M.~Brehmer and T~Munzner.
\newblock A multi-level typology of abstract visualization tasks.
\newblock In {\em Infovis 2013}, pages 2376--2385, 2013.

\bibitem{IV_Collins}
C.~Collins, G.~Penn, and Carpendale S.
\newblock Bubble sets: Revealing set relations with isocontours over existing
  visualizations.
\newblock In {\em Visualization and Computer Graphics}, pages 1009--1016, 2009.

\bibitem{fabrikant06}
S.I. Fabrikant, D.R. Monteilo, and David~M. Mark.
\newblock The distance-similarity metaphor in region-display spatializations.
\newblock {\em Computer Graphics and Applications, IEEE}, 26(4):34--44, 2006.

\bibitem{IV_Farrugia}
M.~Farrugia and A.~Quigley.
\newblock Effective temporal graph layout: a comparative study of animation
  versus static display methods.
\newblock {\em Information Visualization}, 10(1):47--64, 2011.

\bibitem{mocs}
Daniel Fried and Stephen~G. Kobourov.
\newblock Maps of computer science.
\newblock In {\em 7th IEEE PacificVis Symposium}, 2014.
\newblock To appear.

\bibitem{IV_Gansner}
E.~R. Gansner, Y.~Hu, and S.~G. Kobourov.
\newblock Visualizing graphs and clusters as maps.
\newblock In {\em Computer Graphics and Applications}, pages 2259--2267, 2010.

\bibitem{IV_ghani2}
S.~Ghani, N.~Elmqvist, and J.~S. Yi.
\newblock Perception of animated node-link diagrams for dynamic graphs.
\newblock {\em Computer Graphics Forum}, 31(1):1205--1214, 2012.

\bibitem{Ghani2011}
Sohaib Ghani and Niklas Elmqvist.
\newblock Improving revisitation in graphs through static spatial features.
\newblock In {\em Proceedings of Graphics Interface 2011}, GI '11, pages
  175--182, 2011.

\bibitem{IV_Huang1}
W.~Huang and P.~Eades.
\newblock How people read graphs.
\newblock In {\em Asia-Pacific symposium on Information visualisation}, pages
  51--58, 2005.

\bibitem{IV_Huang2}
W.~Huang, P.~Eades, and S.~H. Hong.
\newblock Beyond time and error: a cognitive approach to the evaluation of
  graph drawings.
\newblock In {\em conference on BEyond time and errors: novel evaLuation
  methods for Information Visualization}, 2008.

\bibitem{IV_Huang3}
W.~Huang, P.~Eades, and S.~H. Hong.
\newblock Measuring effectiveness of graph visualizations: A cognitive load
  perspective.
\newblock In {\em Information Visualization}, pages 139--152, 2014.

\bibitem{IV_Huang4}
W.~Huang, S.~H. Hong, and P.~Eades.
\newblock Predicting graph reading performance: a cognitive approach.
\newblock In {\em Asia-Pacific Symposium on Information Visualisation}, pages
  207--216, 2006.

\bibitem{IV_Radu}
R.~Jianu, A.~Rusu, Y.~Hu, and D.~Taggart.
\newblock How to display group information on node-link diagrams: an
  evaluation.
\newblock In {\em IEEE Trans. Visualization and Computer Graphics (TVCG)},
  2014.

\bibitem{IV_BL}
B.~Lee, C.~Plaisant, C.~Parr, J.~D. Fekete, , and N.~Henry.
\newblock Task taxonomy for graph visualization.
\newblock In {\em the 2006 AVI Workshop on Beyond Time and Errors: Novel
  Evaluation Methods For information Visualization (BELIV '06)}, pages 81--85,
  2006.

\bibitem{IV_Mishra}
N.~Mishra, R.~Schreiber, Stanton. I., and R.~E. Tarjan.
\newblock Clustering social network.
\newblock In {\em Workshop on Algorithms and Models for the Web-Graph
  (WAW2007)}, pages 56--67, 2007.

\bibitem{Graph_layout_evaluation_Purchase}
Helen~C. Purchase.
\newblock Which aesthetic has the greatest effect on human understanding?
\newblock In Giuseppe Di~Battista, editor, {\em International Symposium on
  Graph Drawing (GD97)}, volume 1353 of {\em Lecture Notes in Computer
  Science}, pages 248--261. Springer, 1997.

\bibitem{Graph_layout_evaluation_UML}
Helen~C. Purchase, David~A. Carrington, and Jo-Anne Allder.
\newblock Empirical evaluation of aesthetics-based graph layout.
\newblock {\em Empirical Software Engineering}, 7(3):233--255, 2002.

\bibitem{Sk02}
Andr\'e Skupin.
\newblock A cartographic approach to visualizing conference abstracts.
\newblock {\em {IEEE} Computer Graphics and Applications}, 22(1):50--58, 2002.

\bibitem{sf-sm-03}
André Skupin and Sara~Irina Fabrikant.
\newblock Spatialization methods: a cartographic research agenda for
  non-geographic information visualization.
\newblock {\em Cartography and Geographic Information Science}, 30:95--119,
  2003.

\end{thebibliography}

\end{document}